\documentclass[doublecol,linenumbers]{epl2} 

\usepackage{subfigure}
\usepackage{amsmath}
\usepackage{amssymb}

\title{Dynamic Behavior of Interacting between Epidemics and Cascades on Heterogeneous Networks
}
\shorttitle{Epidemics and Cascade Dynamics on Heterogeneous Networks} 

\author{ Lurong Jiang\thanks{E-mail: \email{jianglurong@zju.edu.cn}} \and
Xinyu Jin\thanks{E-mail: \email{jinxy@zju.edu.cn}} \and
Yongxiang Xia\thanks{E-mail: \email{xiayx@zju.edu.cn}} \and
Bo Ouyang  \and
Duanpo Wu}
\shortauthor{Lurong Jiang \etal}

\institute{
  Department of Information Science and Electronic Engineering - Zhejiang University, Hangzhou 310027, China
}

\abstract{
Epidemic spreading and cascading failure are two important dynamical processes over complex networks. They have been investigated separately for a long history. But in the real world, these two dynamics sometimes may interact with each other. In this paper, we explore a model combined with SIR epidemic spreading model and local loads sharing cascading failure model. There exists a critical value of tolerance parameter that  whether the epidemic with high infection probability can spread out and infect a fraction of the network in this model. When the tolerance parameter is smaller than the critical value, cascading failure cuts off abundant of paths and blocks the spreading of epidemic locally. While the tolerance parameter is larger than the critical value, epidemic spreads out and infects a fraction of the network. A method for estimating the critical value is proposed. In simulation, we verify the effectiveness of this method in Barab\'asi-Albert (BA) networks. 
}

\begin{document}

\maketitle

\section{Introduction}
\label{sec:intro}

Epidemic spreading and cascading failure are extensively investigated in the research of complex networks. Study of epidemics on networks is an area of special interest including the spreading of infectious diseases, rumors \cite{zhou2007influence} and computer viruses \cite{pastor2001epidemic}, and has deepened our insights into the interplay between topologies and the spreading dynamics \cite{liu2005epidemic,del2013epidemic,guo2013epidemic,hernandez2013epidemic,byungjoon2013suppression}. Cascading failure is another phenomenon that random failures or intentional attacks lead to severe chain reactions propagating through links in networks. It is widely found in power transmission \cite{sachtjen2000disturbances,kinney2005modeling}, communication \cite{cohen2000resilience}, economic \cite{huang2013cascading}, and biological \cite{borrvall2000biodiversity} networks.

Traditionally, these two dynamics correspond to independent topics in the research of complex networks, \textit{i.e.}, when one of them is studied, the other is considered to be irrelevant. But in practice, there are many cases in which they interact with each other and neither of their effects can be ignored. For example, consider a digital virus spreading over a data communication network. In this scenario, an infected router may not deliver data packets normally and cause overloading of nearby routers. These routers then redistribute their loads to other routers. Eventually, a large cascading failure occurs. In turn, the cascading failure also influences the spreading of the virus, because failed nodes cannot deliver data properly, especially cannot deliver copies of the virus properly. A real-world case that computer viruses CODE RED and Nima caused widely reported BGP storms, which is a cascading failure, was observed in September of 2001 \cite{coffman2002network}.

In our previous work \cite{ouyangbo2014epl}, we proposed a model to study the interplay between epidemic spreading and cascading failure. In this model, when the two dynamical processes stop at equilibrium, the nodes both uninfected and un-failed form several clusters. We considered the relative size $G$ of the largest one, \textit{i.e.} the giant component, and found that in both Erd\"os-R\'enyi (ER) random networks and BA scale free networks, with an infectivity over some threshold, a giant component forms only if the tolerance parameter $\alpha$, which captures the capacities of nodes, is within some interval $(\alpha_l,\alpha_u)$. In Ref. \cite{ouyangbo2014epl}, we have stated the reason for this -- with $\alpha$ under $\alpha_l$, a large-scale cascade occurs and almost all nodes fail, while with $\alpha$ over $\alpha_u$, epidemics and cascading failure together kill almost all nodes.

In this paper, we explore this model from the perspective of epidemic spreading, while our previous work inspected it from the perspective of cascading failure. We consider the situation when the infection probability is sufficiently large. In this case, when $\alpha$ is extremely small, the epidemic dies out locally since cascading failure cuts off almost all the paths in the network. When $\alpha$ is sufficiently large, nodes will never be overloaded,  epidemic thus can spreads out. Therefore, there exists a critical value $\alpha_c$ that whether the epidemic with high infection probability can spread out. The critical value $\alpha_c$ is different from $\alpha_u$ in our previous work.  The $\alpha_c$ indicates the critical moment that epidemic starts to spread out but not die out locally, while  $\alpha_u$ implies the critical moment that epidemic and cascading failure kill all nodes in network together. In the rest of this paper, we devote our effort to further studying the interplay between epidemics and cascades, and pay special attention to the critical value $\alpha_c$.

\section{Model}
\label{sec:model}

In Ref. \cite{ouyangbo2014epl}, we link the epidemiological SIR model \cite{kermark1927contributions,heesterbeek2000mathematical} with a local load sharing cascading failure model together to form a hybrid model.

In the SIR model, nodes are in one of the three states: susceptible (S), infected (I), and removed (R). At every time step, the susceptible node becomes infected with probability $\beta$ for every link connecting an infected node with a susceptible node. The infected node becomes removed with probability $\gamma$. The effective spreading rate is $\lambda = \beta / \gamma$. For simplicity, we let $\gamma=1$.

In our model, \textit{removed} nodes in epidemic spreading are assumed to be not functioning and cannot manipulate loads. So the spreading of epidemic will lead to load redistribution, which may cause cascading failure in the network. We assume that the \textit{failed} nodes in cascading failure are not functioning either, so they cannot infect or be infected by others. Because both removed and failed nodes no longer interact with other nodes in the consequent process, they are said to be \textit{inactive}. All other nodes are said to be \textit{active}.

Inspired by previous works \cite{dobson2005loading,sansavini2009deterministic}, we assume that when a node fails, a fixed positive load $\Delta$ is transferred to each of its active neighbors. Overloaded nodes are those whose load exceeds their capacity. It is natural to assume that the capacity $C_v$ of a node $v$ is proportional to its initial load $L_v$ \cite{motter2002cascade,wang2007high,li2008limited,lehmann2010stochastic}
\begin{equation}\label{eq:cap}
C_v=(1+\alpha)L_v,
\end{equation}
where the constant $\alpha$ is the tolerance parameter.

The numerical simulation of the hybrid process in an uncorrelated network with $N$ nodes is summarized as follows:
\begin{enumerate}
\item The initial load of each node is randomly generated according to a uniform distribution on the interval [$L_{min}$,$L_{max}$] and its capacity is defined by Eq. (\ref{eq:cap}). Without loss of generality, we let $L_{min}=0$ and $L_{max}=1$.
\item Randomly select a few nodes in the network as the initial infected nodes.
\item\label{item:epidemic} For every link connecting an infected node with a susceptible node, the susceptible node becomes infected with probability $\lambda$, and for every infected node, it becomes removed with probability 1.
\item\label{item:cascade} The removal may cause load redistribution and possibly cascading failure. Repeat the cascades until there is no overloaded node in the network.
\item If there exist infected nodes in the network, go back to \ref{item:epidemic} to start another step of epidemic spreading; otherwise, the process halts.
\end{enumerate}

Note that we make another assumption that the time scale of cascading failure is much smaller than that of epidemic spreading, so every single step of epidemic spreading is executed until there are no overloaded nodes in the network.

\section{Analysis}
\label{sec:analysis}

 \begin{figure}
\centering
\subfigure[ER random network.]{
\includegraphics[width=0.35\textwidth]{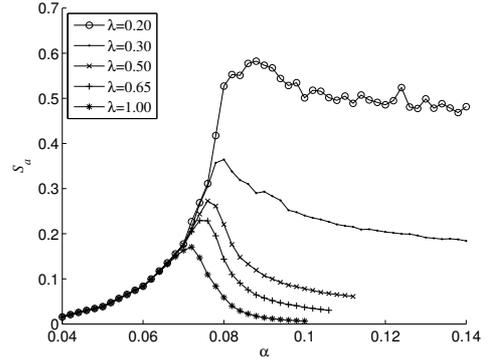}
\label{fig:an_er}
}
\subfigure[BA scale-free network.]{
\includegraphics[width=0.35\textwidth]{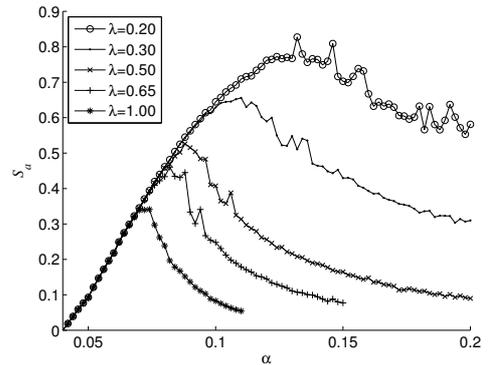}
\label{fig:an_ba}
}
\caption{The fraction of active nodes $S_a$ as a function of tolerance parameter $\alpha$ with varying effective spreading rate  $\lambda$ when the dynamics are terminated. The number of nodes $N$=5000, and the average degree $\langle k \rangle =8$. Each data point is the average over 50 network realizations. In the simulation, $\Delta$, the number of units of load that a failed node shares with each of its neighbors, is set to 0.01.}
\label{fig:an}
\end{figure}

When the process ends at equilibrium, the active nodes in the network form several clusters. In Ref. \cite{ouyangbo2014epl}, we use the fraction $G$ of the largest cluster, namely, the giant component, to measure the performance of the network. But here, for now, we examine the fraction $S_a$ of the active nodes rather than the largest cluster formed by them, because it is closely related to $G$ and more likely to be theoretically tractable. Fig. \ref{fig:an} shows $S_a$ as a function of the tolerance parameter $\alpha$ in ER and BA networks. When $\alpha$ is extremely small, cascading failure kills almost all nodes and epidemic dies out locally, thus $S_a$ is very close to zero. With $\alpha$ increases, the severity of cascading failure reduces and $S_a$ increases. When $\alpha$ is over some critical value, $\alpha_c$, the epidemic starts to spread out and infect a large number of nodes, making $S_a$ decrease. 

\subsection{Preliminaries}

Before we introduce the method to solve the critical value $\alpha_c$ in our model, we present some preliminary remarks.

\begin{figure}
\centering
\includegraphics[scale=0.6]{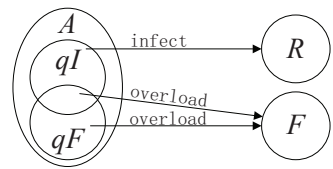}
\caption{The transitions between states. The states $A$, $R$, $F$, $qF$ and $qI$ are abbreviations for active, removed, failed, quasi-failed and quasi-infected states respectively. For an active node, if at least one of its edges participates in bond percolation process, it can be called quasi-infected node, and it will come to be a removed or failed node. While an active node takes part in site percolation process, it is described as quasi-failed node, and finally it will turn out to be a failed node explicitly. }
\label{fig:StatesTransitionDiagram}
\end{figure}

The spread of disease is equivalent to a \textit{bond percolation} process \cite{PhysRevE.66.016128,PhysRevE.82.016101}. With effective spreading rate $\lambda$, the equivalent process is as follows: select an edge from network with probability $\lambda$, and the outbreak of disease means that the set of chosen edges forms a connectivity cluster. The cascading failure is a \textit{site percolation} process \cite{PhysRevE.77.046117,PhysRevE.83.056107}. There are several nodes participating in both bond percolation and site percolation processes. We assume that these nodes belong to site percolation process rather than bond percolation process. The reason for this assumption is that, although the cascading failure is triggered by the spreading of epidemic, once cascade starts with a small value of $\alpha$, most of the paths linked to infected nodes are cut off by cascading failure, and thus the number of failed nodes from cascading failure will be much larger than newly infected nodes. Therefore, most of the nodes participating in both bond percolation and site percolation processes are considered to be failed in cascades eventually, rather than removed in epidemic spreading. Note that there is a premise that $\alpha$ is small, which means most of the inactive nodes come from cascades while epidemic is restricted in a small scale, since the time scale of cascading failure is much smaller than that of epidemic spreading. Based on the assumption above, we define two types of excessive states to describe the dynamics before a node becomes inactive as shown in Fig. \ref{fig:StatesTransitionDiagram}:

\begin{quote}
\item\label{item:quasi-infected} \textit{Quasi-infected} node, a node at least one of whose edges participates in bond percolation process.

\item\label{item:quasi-failed} \textit{Quasi-failed} node, a node takes part in site percolation process.
\end{quote}

A quasi-failed node will fail in cascades undoubtedly, while quasi-infected node will become removed or failed in the end. Denote $s_k$ as the probability that a node of degree $k$ happens to be a quasi-failed node, and $b_k$ that it happens to be a quasi-infected node. Eventually, this node happens to be failed with probability $s_k$, or be removed with probability $(1-s_k)b_k$. 

Further more, in this model, we call a node:
\begin{quote}
An \textit{$m$-vulnerable} node, if and only if it has $m$ quasi-failed or quasi-infected neighbors, and it becomes inactive if all of the $m$ neighbors fail and share load with it.

A \textit{Vulnerable} node, if it is an $m$-vulnerable node, where $m \geq 0$.
\end{quote}

The probability that a node of degree $k$ happens to be a vulnerable node is

\begin{equation}
\label{eq:fragile_def}
f_k = \sum_{m=0}^{k} f_k^{(m)} 					,
\end{equation}

\noindent where $f_k^{(m)}$ is the probability that it is an $m$-vulnerable node.

When $\alpha$ is small, the number of quasi-failed nodes is larger than that of quasi-infected nodes, as the time scale of cascading failure is much smaller than that of epidemic spreading. Thus vulnerable nodes can be approximated to quasi-failed nodes, as $f_k \approx s_k$.

\subsection{Critical Value of Tolerance Parameter }

In an uncorrelated network,  consider a node $v$ of degree $k$ and any of its neighbor $w$. The probability of node $w$ with degree $j$ is $\frac{j p_j}{\sum_i i p_i}=\frac{j p_j}{\langle k \rangle}$. Node $w$ becomes vulnerable  or infected  with probability $1-(1-s_j)(1-b_j)=s_j+(1-s_j)b_j$. Therefore, the probability of node $v$'s arbitrary neighbor being quasi-failed or quasi-infected is
\begin{equation}
\label{eq:sigmaFandR}
\sum_j \frac{j p_j (s_j + (1-s_j)b_j)}{\langle k \rangle} \triangleq \sigma_f + \sigma_r 
\end{equation}

\noindent where $\sigma_f$ and $\sigma_r$ are the probabilities that node $v$'s arbitrary neighbor happens to be a failed or removed node at equilibrium, respectively, as follows, 

\begin{align}
\sigma_f \triangleq &\sum_{j}\frac{j p_j s_j}{\langle k \rangle} \label{eq:sigma_f_def} \;  \\
\sigma_r \triangleq &\sum_{j}\frac{j p_j (1-s_j)b_j}{\langle k \rangle} \label{eq:sigma_r_def} \; 
\end{align}

According to Eq. (\ref{eq:sigmaFandR}), the probability that exactly $m$ of node $v$'s neighbors are quasi-failed is

\begin{equation}
\binom{k}{m} \left(\sigma_f + \sigma_r\right)^m \left(1 - \sigma_f - \sigma_r\right)^{k-m} 		
\end{equation}

For node $v$, its initial load $L_v$ is uniformly distributed in the range [0,1], so the cumulative distribution function of $L_v$ is
\begin{equation}
\label{eq:load_cdf}
P\{ L_v<l \} \triangleq \varphi(l) =
\begin{cases}
0 	& \ \text{$l \leq 0$} \\
l  	& \ \text{$0 < l \leq 1$} \\
1 	& \ \text{$l > 1$}
\end{cases} \; 		.
\end{equation}

The probability that node $v$'s failure is triggered by the failure and loads sharing from exact $m$ of its neighbors is 

\begin{equation}
P\!\left\{ L_v\!+\!m\!\Delta>(1\!+\!\alpha)L_v \right\}=P\!\left\{ L_v<\frac{m \Delta}{\alpha} \right\} = \varphi\!\left( \frac{m \Delta}{\alpha} \right)		.
\end{equation}

Then, the probability that node $v$ is an $m$-vulnerable node is

\begin{equation}
\label{eq:f_km}
f_k^{(m)}=\binom{k}{m}\!\left(\!\sigma_f\!+\!\sigma_r\!\right)\!^m\left(1\!-\!\left(\!\sigma_f\!+\!\sigma_r\!\right)\!\right)^{k\!-\!m}\varphi\left(\frac{m\Delta}{\alpha}\right).
\end{equation}

We substitute Eq. (\ref{eq:f_km}) into Eq. (\ref{eq:fragile_def}), and get

\begin{small}
\begin{equation}
\label{eq:site_percolation_prob}
\begin{split}
&s_k=f_k=\sum\limits_{m=0}^{k}\binom{k}{m}\!\left(\!\sigma_f\!+\!\sigma_r\!\right)\!^m\!\left(\!1\!-\!\sigma_f\!-\!\sigma_r\!\right)\!^{k\!-\!m}\varphi\!\left(\!\frac{m\Delta}{\alpha}\!\right)\! \\
&=
\begin{cases}
\!\frac{k\!\left(\!\sigma_f\!+\!\sigma_r\!\right)\!\Delta}{\alpha},  \hfill \text{$k \leq \left[\frac{\alpha}{\Delta}\right]$}	\\ \\	
\!1\!-\!\sum\limits_{m=0}^{\left[\!\alpha/\Delta\!\right]}\!\binom{k}{m}\!\left(\!\sigma_f\!+\!\sigma_r\!\right)\!^m\!\left(\!1\!-\!\sigma_f\!-\!\sigma_r\!\right)\!^{k\!-\!m}\!\left(\!1\!-\!\frac{m\Delta}{\alpha}\!\right)\!,	
\hfill \text{$k > \left[\frac{\alpha}{\Delta}\right]$}
\end{cases}
\end{split}
\end{equation}
\end{small}

\noindent where $\left[ \bullet \right]$ is the floor function.

Eq. (\ref{eq:site_percolation_prob}) provides the probability that arbitrary node in network turns out to be a quasi-failed node. Then, we will handle the probability of a node comes to be a quasi-infected node  $b_k$. We know the probability that node $v$ of degree $k$ survives with active state at equilibrium is $(1-s_k)(1-b_k)$. These nodes are active nodes, that they are neither quasi-failed nodes nor quasi-infected nodes. The state of these active nodes is determined by their neighbors:

\begin{enumerate}
\item\label{itm:neib_case1}
If their neighbors are quasi-failed or quasi-infected and finally become failed in cascades, then these active nodes may turn out to be failed as the overload caused by their neighbors. The probability of this situation happens is $\sum_{j}\frac{j p_j \left(s_j + (1-s_j)b_j(1-\lambda)\right)}{\langle k \rangle}=\sigma_f+(1-\lambda)\sigma_r$.

\item If their neighbors are active nodes, i.e., the neighbors  are neither quasi-failed nor quasi-infected nodes, then these active nodes will not fail. The probability of this situation happens is $\sum_{j}\frac{j p_j (1-s_j)(1-b_j)}{\langle k \rangle}=1-\!\left(\!\sigma_f\!+\!\sigma_r\!\right)\!$.
\end{enumerate}

Thus the probability that a node survives with active state, $h_k$, can be written follows its neighbors states as

\begin{small}
\begin{equation}
\label{eq:h_k}
\begin{split}
&h_k=
\sum_{m=0}^{k}\binom{k}{m}\!\left(\!1\!-\!\varphi\!\left(\!\frac{m\Delta}{\alpha}\!\right)\!\right)\!\left(\!\sigma_f\!+\!(1\!-\!\lambda)\sigma_r\!\right)\!^m\!\left(\!1\!-\!\sigma_f\!-\!\sigma_r\!\right)\!^{k\!-\!m} \\
=&
\begin{cases}
\!\left(\!1\!-\!\lambda\sigma_r\!\right)\!^k\!-\!\frac{k\!\left(\!\sigma_f\!+\!(1\!-\!\lambda)\sigma_r\!\right)\!\Delta}{\alpha}\!\left(\!1\!-\!\lambda\sigma_r\!\right)\!^{k\!-\!1}, \hfill \text{$k \leq \left[\frac{\alpha}{\Delta}\right]$} 	\\ \\		
\sum\limits_{\!m\!=\!0\!}^{\!\left[\!\alpha\!/\!\Delta\!\right]\!}\binom{k}{m}\!\left(\!\sigma_f\!+\!(1\!-\!\lambda)\sigma_r\!\right)\!^m\!\left(\!1\!-\!\sigma_f\!-\!\sigma_r\!\right)\!^{k\!-\!m}\!\left(\!1\!-\!\frac{m\Delta}{\alpha}\!\right)\!,
\hfill \text{$k > \left[\frac{\alpha}{\Delta}\right]$}			
\end{cases}
\end{split}
\end{equation}
\end{small}

In the first line of Eq. (\ref{eq:h_k}), $\left(1-\varphi\left(\frac{m\Delta}{\alpha}\right)\right)$
indicates the situation that $m$ of node $v$'s neighbors are quasi-failed or quasi-infected nodes but it is still active. Based on the discussion above, we have $h_k = (1-s_k)(1-b_k) = 1-s_k-(1-s_k)b_k$, where $s_k$ is given in Eq. (\ref{eq:site_percolation_prob}). According to  Eq. (\ref{eq:site_percolation_prob}) and Eq. (\ref{eq:h_k}), we can get

\begin{small}
\begin{equation}
\begin{split}
\label{eq:bond_percolation_prob}
&(1\!-\!s_k)b_k = r_k \\
\!=\!&
\begin{cases}
\!1\!-\!\frac{k\!\left(\!\sigma_f\!+\!\sigma_r\!\right)\!\Delta}{\alpha}\!-\!\left(\!1\!-\!\lambda\sigma_r\!\right)\!^k\!+\!\frac{k\!\left(\!\sigma_f\!+\!(1\!-\!\lambda)\sigma_r\!\right)\!\Delta}{\alpha}\!\left(\!1\!-\!\lambda\sigma_r\!\right)\!^{k\!-\!1} , \\
\hfill \text{$k \leq \left[\frac{\alpha}{\Delta}\right]$} \\ \\
\!\sum\limits_{\!m\!=\!0\!}^{\!\left[\!\alpha\!/\!\Delta\!\right]\!}\binom{\!k\!}{\!m\!}\!\left(\!\left(\!\sigma\!_f\!+\!\sigma\!_r\!\right)\!^m\!-\!\left(\!\sigma\!_f\!+\!\left(\!1\!-\!\lambda\!\right)\!\sigma\!_r\!\right)\!^m\!\right)\!\left(\!1\!-\!\sigma\!_f\!-\!\sigma\!_r\!\right)\!^{k\!-\!m\!}\!\left(\!1\!-\!\frac{\!m\!\Delta\!}{\!\alpha\!}\!\right)\!, \\
\hfill \text{$k > \left[\frac{\alpha}{\Delta}\right]$}
\end{cases}
\end{split}
\end{equation}
\end{small}

\noindent where $r_k$ is the probability that a node of degree $k$ be a removed node in epidemic spreading finally. 

Now we complete the derivation by Eq. (\ref{eq:sigma_f_def}), (\ref{eq:sigma_r_def}), (\ref{eq:site_percolation_prob}), (\ref{eq:bond_percolation_prob}). These four equations depend on each other. We can get $f_k$, $r_k$, the fraction of removed nodes $S_r$, and the fraction of failed nodes $S_f$ at equilibrium from these four equations by the following iteration process:

\begin{enumerate}
\item  Set the initial values. Assign a very small initial value to $r_{k_{min}}$, where $k_{min}$ represents the minimum degree of all nodes. Let $r_k=0 (k > k_{min})$ and $f_k = 0$. Substitute $r_k$ and $f_k$ into Eq. (\ref{eq:sigma_f_def}) and Eq. (\ref{eq:sigma_r_def}) to get the initial values of $\sigma_f$ and $\sigma_r$.

\item Substitute  $\sigma_f$ and $\sigma_r$ into Eq. (\ref{eq:site_percolation_prob}) and Eq. (\ref{eq:bond_percolation_prob}) to get the values of $r_k$ and $f_k$. Substitute the new values of $r_k$ and $f_k$ into Eq. (\ref{eq:sigma_f_def}) and Eq. (\ref{eq:sigma_r_def}), update $\sigma_f$ and $\sigma_r$. Repeat this step until convergence.

\item  Get the fraction of removed and failed nodes at equilibrium by $S_r = \sum_k p_k r_k $ and $S_f = \sum_k p_k f_k $ respectively.  With $\alpha$ increases, the $S_r$ will change from zero to a positive value at $\alpha_c$ and this indicates when the epidemic starts to spread out.
\end{enumerate}

\section{Simulations}
\label{sec:simulation}
We mainly focus on the fraction of active nodes in simulation, and validate the theoretical results of critical value $\alpha_c$. We carry out the simulations in BA networks with 5000 nodes and varied average degrees $\langle k \rangle=8,10$. The results are averaged over 50 realizations in each of which the network is independently generated.  The unit of load shared in cascading failure, $\Delta$, is fixed at $0.01$. 

\begin{figure}[!htbp]
\centering
\subfigure[]{
\includegraphics[width=0.4\textwidth]{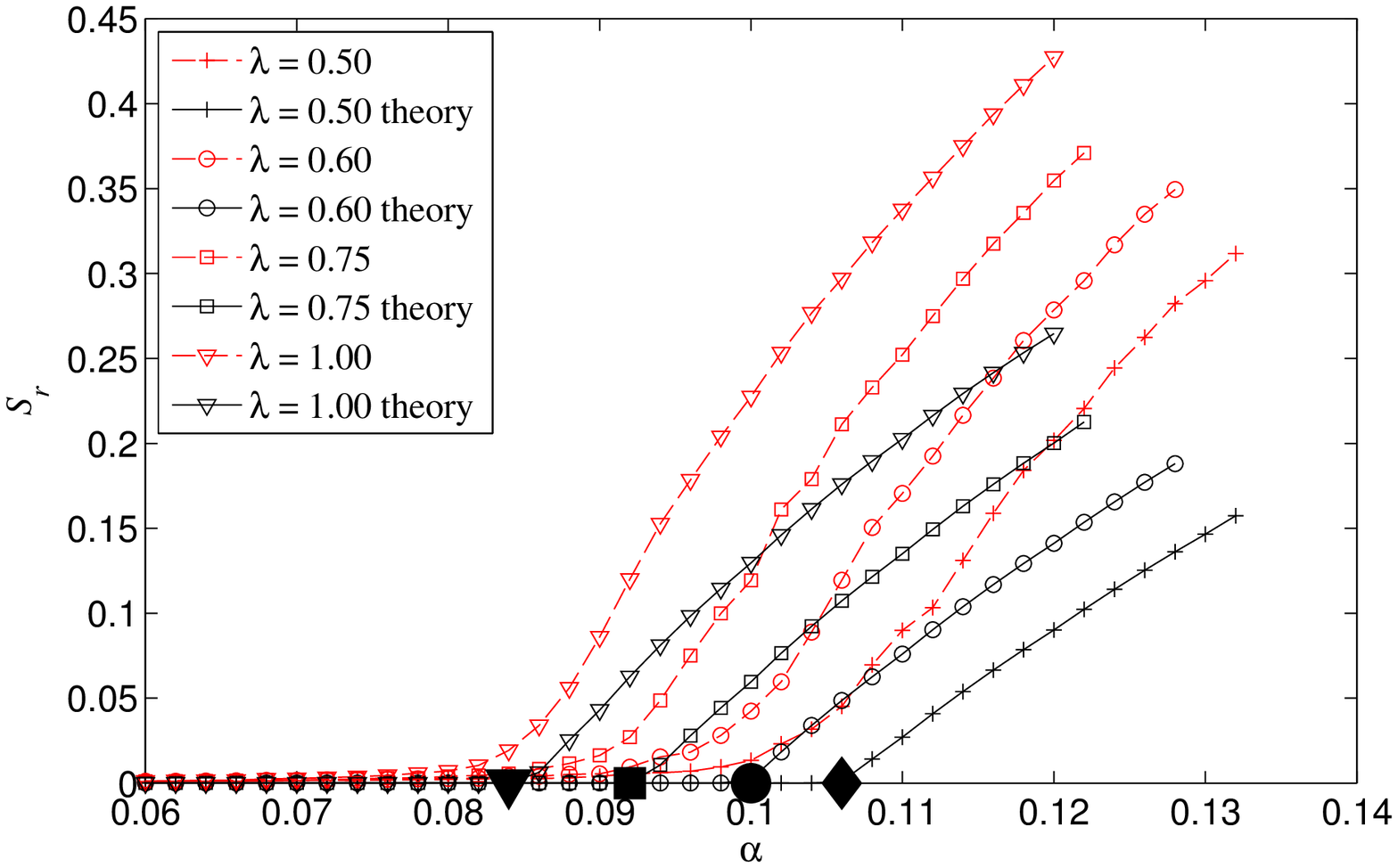}
\label{fig:sigma_r_lambda}
}
\subfigure[]{
\includegraphics[width=0.4\textwidth]{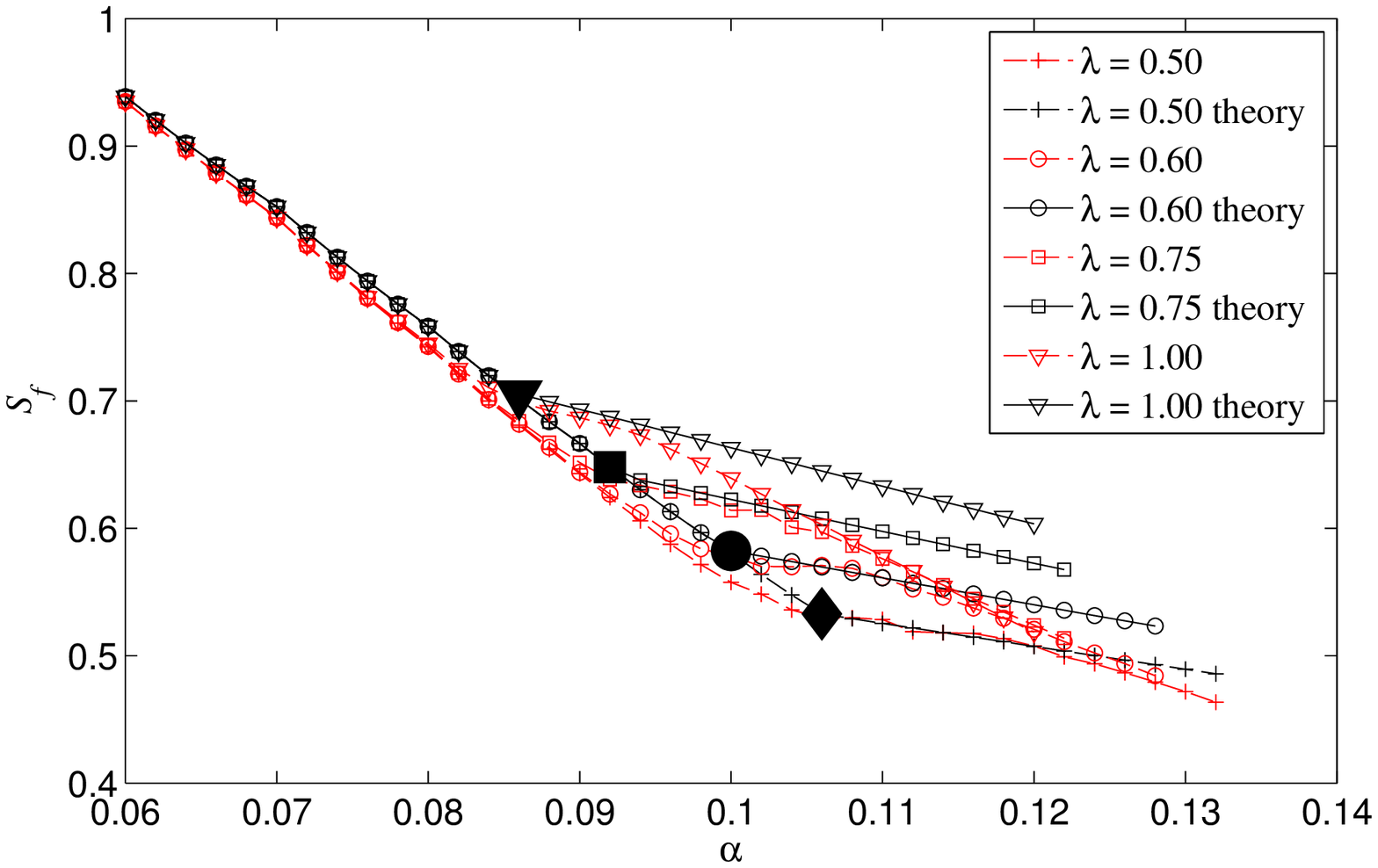}
\label{fig:sigma_f_lambda}
}
\caption{(Color online) The fractions of removed (a) and failed (b) nodes after the dynamics are terminated in BA networks with $\langle k \rangle = 10$ and varied $\lambda$ values are displayed. The red dashed lines are simulation results and the black solid lines are theoretical results. For theoretical results, the turning points (whose abscissas are $\alpha_c$) with $\lambda = 0.5, 0.6, 0.75, 1$ are marked by solid diamond ($\blacklozenge$), circle ($\bullet$), square ($\blacksquare$), and down-pointing triangle ($\blacktriangledown$) respectively. }
\label{fig:sigma_r_and_f_lambda}
\end{figure}

Before exploring the fraction of active nodes, we firstly investigate the complementary states consisting of failed and removed nodes. Fig. \ref{fig:sigma_r_and_f_lambda} presents the fractions of failed nodes $S_f$ and removed nodes $S_r$  with varied effective spreading rate  $\lambda$ in BA networks. When $\alpha$ is small , the theoretical results are in good agreement with simulation results. While $\alpha$ is large, there is a deviation between theoretical result and simulation result. The reason for this deviation is that nodes participates both bond percolation (epidemic spreading) and site percolation processes (cascading failure) are suppose to be the latter one in theoretical prediction. In simulation, a considerable amount of nodes are infected and become removed when $\alpha$ is large. We can easily find that each of the curves of simulation and theoretical results has a turning point at abscissa $\alpha_c$. In Fig. \ref{fig:sigma_r_lambda}, when $\alpha$ is smaller than $\alpha_c$, $S_r$ is very close to zero, which means the epidemic dies out locally. While $\alpha$ is larger than $\alpha_c$, $S_r$ is larger than zero, indicating the epidemic spreads out and infects a portion of the network. In Fig. \ref{fig:sigma_f_lambda}, when $\alpha$ is smaller than $\alpha_c$, $S_f$ decreases with increasing of $\alpha$. While $\alpha$ is larger than $\alpha_c$, $S_f$ decreases more slowly, indicating the cascading failure slows down, and epidemic spreads out and infects a fraction of the network. This is an interesting phenomenon that even if the infection probability is very large, the epidemic still dies out when $\alpha$ is smaller than $\alpha_c$.
This is because, when $\alpha$ is small, very few initial infected nodes can lead to severe cascading failure, which cuts off the spreading paths of the epidemic. 

\begin{figure}[!htbp]
\centering
\subfigure[$\lambda = 0.5$]{
\includegraphics[width=0.22\textwidth]{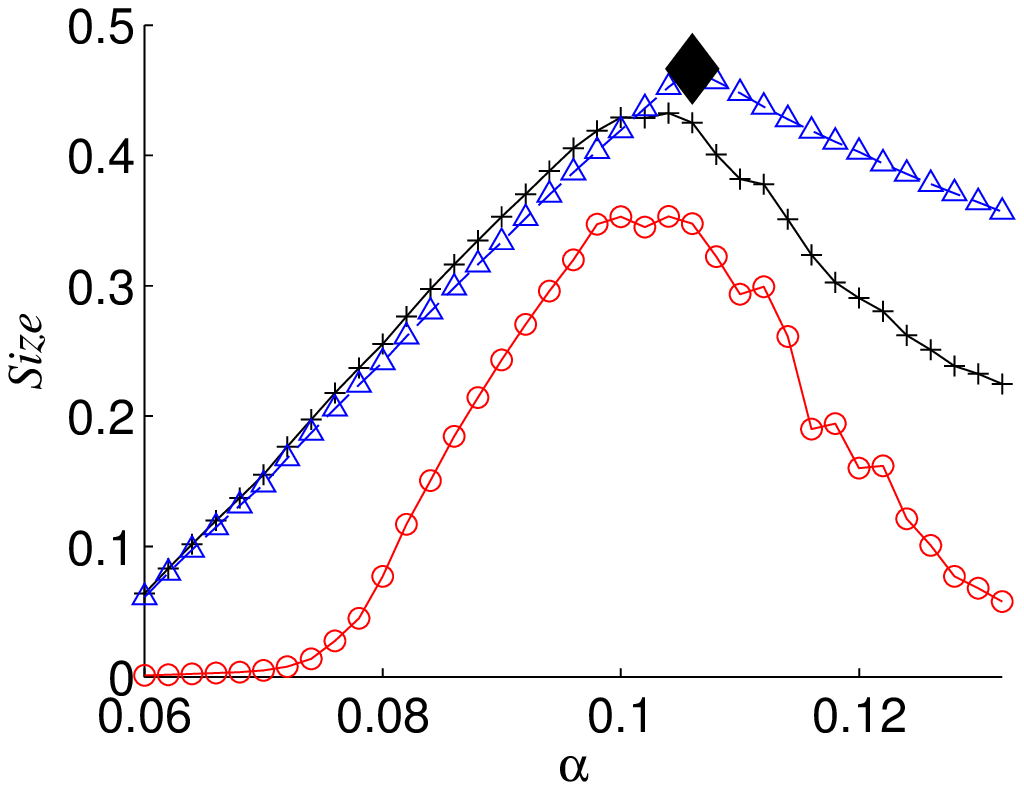}
\label{fig:triCompare_lambda_0.5}
}
\subfigure[$\lambda = 0.6$]{
\includegraphics[width=0.22\textwidth]{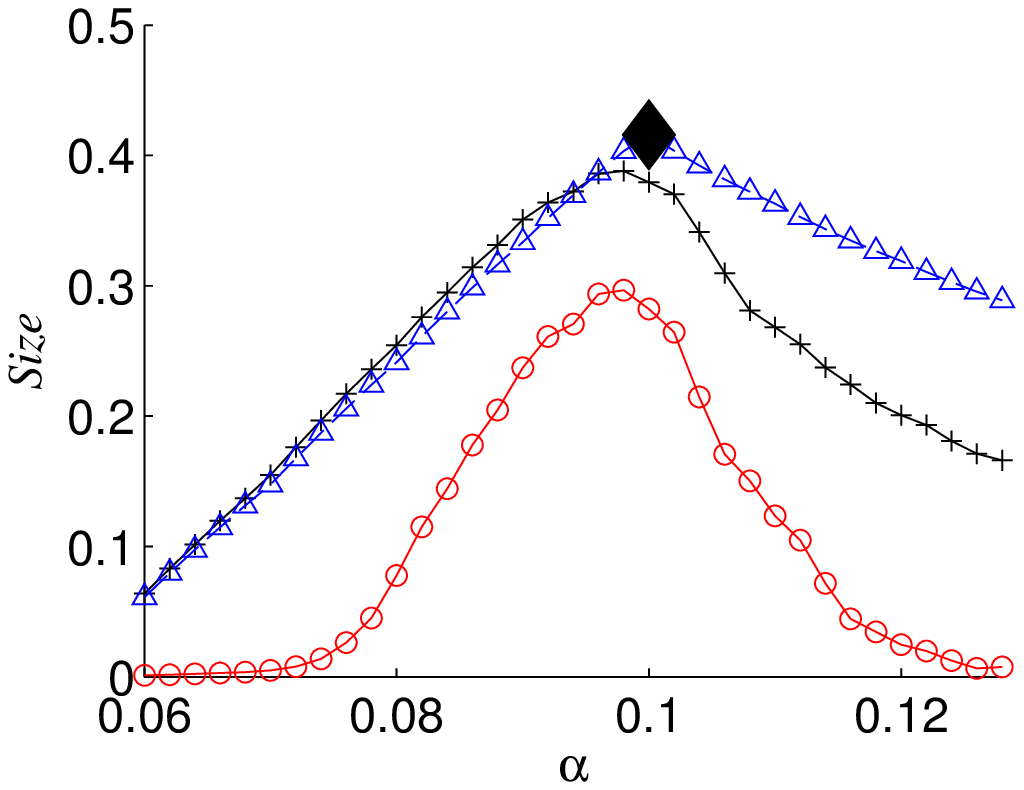}
\label{fig:triCompare_lambda_0.6}
}
\subfigure[$\lambda = 0.75$]{
\includegraphics[width=0.22\textwidth]{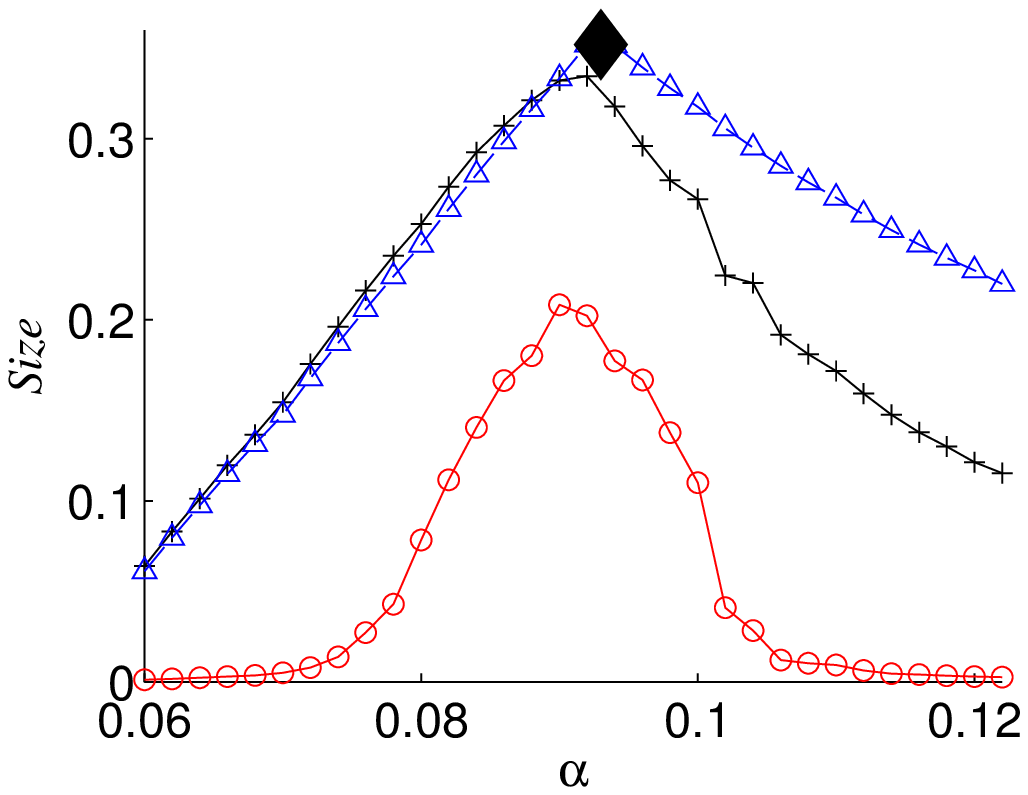}
\label{fig:triCompare_lambda_0.75}
}
\subfigure[$\lambda = 1$]{
\includegraphics[width=0.22\textwidth]{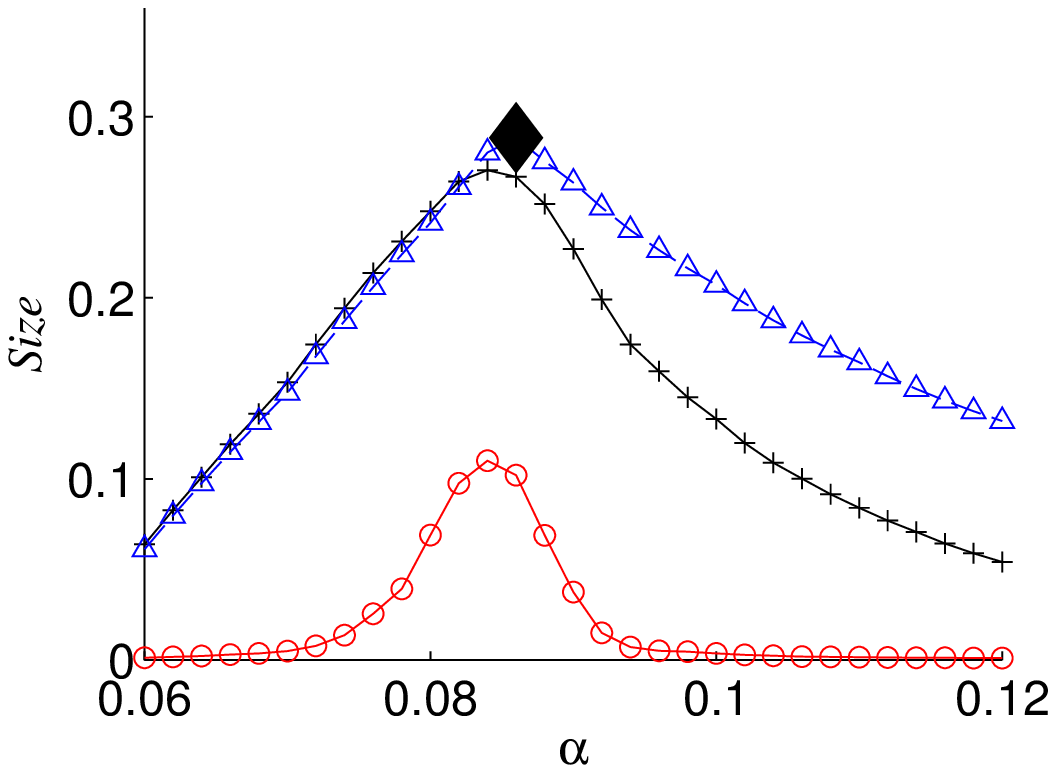}
\label{fig:triCompare_lambda_1}
}
\caption{(Color online) Comparison of the fraction of active nodes $S_a$ and the giant component $G$ after dynamics are terminated in BA networks: theoretical results of all remaining active nodes $S_a$ ({\color{blue}{$\vartriangle$}}), simulation results of $S_a$ ({\color{black}{$+$}}), and giant components of active nodes $G$ ({\color{red}{$\circ$}}). For theoretical results, the turning points (whose abscissas are $\alpha_c$) are marked by solid diamond ($\blacklozenge$).The average degree of BA networks here are $\langle k \rangle = 10$.}
\label{fig:triCompare_lambda}
\end{figure}


Then we explore the fraction of active nodes $S_a$ and the fraction of the giant component $G$ formed by them when the two dynamics of epidemic spreading and cascading failure are terminated, as shown in Fig. \ref{fig:triCompare_lambda}. Simulation and theoretical results of $S_a$ are both plotted. Because in a data communication network, the fraction of the largest connected sub-network formed by active nodes (the giant component of active nodes, $G$) is the important fraction to capture the robustness of the network with respect to structural damage and maintain the network functions \cite{PhysRevLett.106.048701}. Though the sizes of $G$ and $S_a$ are different, their humps meet the peaks at almost the same $\alpha$ value which is very close to the critical value $\alpha_c$. This means that our theoretical method can be used to estimate the critical value $\alpha_c$ as well as the optimal value of $\alpha$ at which most active nodes are remained. It is worth noting that the fraction of active nodes is remarkably larger than that of standard SIR model (the size for remaining active nodes is very close to zero with large $\lambda$ in standard SIR model), because in our model, the cascading failure blocks the spreading of the epidemic, as we have mentioned above.


\section{Conclusion}
\label{sec:conclusion}
In this paper, we explore a model based on the SIR epidemic spreading model and a local load sharing cascading failure model. With change of tolerance parameter $\alpha$,  there exists a critical value $\alpha_c$ that whether the epidemic with large $\lambda$ is bocked by cascading failure. When $\alpha < \alpha_c$, cascading failure cuts off abundant of paths and block the spreading of epidemic locally. While $\alpha > \alpha_c$, epidemic spreads out and infects a fraction of the network. We propose a method for estimating $\alpha_c$ which is applicable in uncorrelated networks. In simulation, we validate the effectiveness of this method in BA networks. 


\bibliographystyle{eplbib}
\bibliography{ref}

\begin{thebibliography}{10}
\expandafter\ifx\csname url\endcsname\relax\def\url#1{\texttt{#1}}\fi

\bibitem{zhou2007influence}
\Name{Zhou J., Liu Z. \and Li B.} \REVIEW{Phys. Lett. A}{368}{2007}{458}.

\bibitem{pastor2001epidemic}
\Name{Pastor-Satorras R. \and Vespignani A.} \REVIEW{Phys. Rev.
  Lett.}{86}{2001}{3200}.

\bibitem{liu2005epidemic}
\Name{Liu Z. \and Hu B.} \REVIEW{Europhys. Lett.}{72}{2005}{315}.

\bibitem{del2013epidemic}
\Name{Del~Genio C.~I. \and House T.} \REVIEW{Phys. Rev. E}{88}{2013}{040801}.

\bibitem{guo2013epidemic}
\Name{Guo D., Trajanovski S., van~de Bovenkamp R., Wang H. \and Van~Mieghem P.}
  \REVIEW{Phys. Rev. E}{88}{2013}{042802}.

\bibitem{hernandez2013epidemic}
\Name{Hern\'andez D.~G. \and Risau-Gusman S.} \REVIEW{Phys. Rev.
  E}{88}{2013}{052801}.

\bibitem{byungjoon2013suppression}
\Name{Min B., Goh K.-I. \and Kim I.-M.} \REVIEW{Europhys.
  Lett.}{103}{2013}{50002}.

\bibitem{sachtjen2000disturbances}
\Name{Sachtjen M., Carreras B. \and Lynch V.} \REVIEW{Phys. Rev.
  E}{61}{2000}{4877}.

\bibitem{kinney2005modeling}
\Name{Kinney R., Crucitti P., Albert R. \and Latora V.} \REVIEW{Eur. Phys. J.
  B}{46}{2005}{101}.

\bibitem{cohen2000resilience}
\Name{Cohen R., Erez K., Ben-Avraham D. \and Havlin S.} \REVIEW{Phys. Rev.
  Lett.}{85}{2000}{4626}.

\bibitem{huang2013cascading}
\Name{Huang X., Vodenska I., Havlin S. \and Stanley H.~E.} \REVIEW{Sci.
  Rep.}{3}{2013}{1219}.

\bibitem{borrvall2000biodiversity}
\Name{Borrvall C., Ebenman B. \and Jonsson T.} \REVIEW{Ecol.
  Lett.}{3}{2000}{131}.

\bibitem{coffman2002network}
\Name{Coffman~Jr E., Ge Z., Misra V. \and Towsley D.} \Book{Network resilience:
  exploring cascading failures within bgp} in proc. of \Book{Proc. 40th Annual
  Allerton Conference on Communications, Computing and Control, IL, USA,} 2002
  pp. 1--10.

\bibitem{ouyangbo2014epl}
\Name{Ouyang B., Jin X., Xia Y., Jiang L. \and Wu D.} \REVIEW{Europhys.
  Lett.}{106}{2014}{28005}.

\bibitem{kermark1927contributions}
\Name{Kermark M. \and Mckendrick A.} \REVIEW{Proc. R. Soc. Lond.
  A}{115}{1927}{700}.

\bibitem{heesterbeek2000mathematical}
\Name{Heesterbeek J.} \Book{Mathematical Epidemiology of Infectious Diseases:
  Model Building, Analysis and Interpretation} Vol.~5 (John Wiley \& Sons)
  2000.

\bibitem{dobson2005loading}
\Name{Dobson I., Carreras B.~A. \and Newman D.~E.} \REVIEW{Probab. Eng. Inf.
  Sci.}{19}{2005}{15}.

\bibitem{sansavini2009deterministic}
\Name{Sansavini G., Hajj M., Puri I. \and Zio E.} \REVIEW{Europhys.
  Lett.}{87}{2009}{48004}.

\bibitem{motter2002cascade}
\Name{Motter A.~E. \and Lai Y.-C.} \REVIEW{Phys. Rev. E}{66}{2002}{065102}.

\bibitem{wang2007high}
\Name{Wang B. \and Kim B.~J.} \REVIEW{Europhys. Lett.}{78}{2007}{48001}.

\bibitem{li2008limited}
\Name{Li P., Wang B.-H., Sun H., Gao P. \and Zhou T.} \REVIEW{Eur. Phys. J.
  B}{62}{2008}{101}.

\bibitem{lehmann2010stochastic}
\Name{Lehmann J. \and Bernasconi J.} \REVIEW{Phys. Rev. E}{81}{2010}{031129}.

\bibitem{PhysRevE.66.016128}
\Name{Newman M. E.~J.} \REVIEW{Phys. Rev. E}{66}{2002}{016128}.

\bibitem{PhysRevE.82.016101}
\Name{Karrer B. \and Newman M. E.~J.} \REVIEW{Phys. Rev. E}{82}{2010}{016101}.

\bibitem{PhysRevE.77.046117}
\Name{Gleeson J.~P.} \REVIEW{Phys. Rev. E}{77}{2008}{046117}.

\bibitem{PhysRevE.83.056107}
\Name{Hackett A., Melnik S. \and Gleeson J.~P.} \REVIEW{Phys. Rev.
  E}{83}{2011}{056107}.

\bibitem{PhysRevLett.106.048701}
\Name{Serrano M.~A., Krioukov D. \and Bogu\~n\'a M.} \REVIEW{Phys. Rev.
  Lett.}{106}{2011}{048701}.

\end{thebibliography}

\end{document}